# Equivalence of two mathematical forms for the bound angular momentum of the electromagnetic field*

## A. M. STEWART

Department of Theoretical Physics,
Research School of Physical Sciences and Engineering,
The Australian National University,
Canberra, ACT 0200, Australia.

**Abstract:**
It is shown that the mathematical form, obtained in a recent paper, for the angular momentum of the electromagnetic field in the vicinity of electric charge is equivalent to another form obtained previously by Cohen-Tannoudji, Dupont-Roc and Gilbert. In this version of the paper an improved derivation is given.

The angular momentum $\mathbf{J}(t)$ of the classical electromagnetic field in terms of the electric $\mathbf{E}(\mathbf{x}, t)$ and magnetic field $\mathbf{B}(\mathbf{x}, t)$ [1]

$$\mathbf{J}(t) = \frac{1}{4\pi c} \int d^3 x \, \mathbf{x} \times [\mathbf{E}(\mathbf{x},t) \times \mathbf{B}(\mathbf{x},t)] \qquad (1)$$

(Gaussian units, bold font denotes a three-vector) has been treated in a paper published recently in this journal [2] by means of the Helmholtz theorem on the decomposition of vector fields. It was shown there that the angular momentum associated with a volume of space far from the influence of electric charge may be expressed as a sum of three terms: a) a volume integral of angular momentum of spin-like character, b) a volume integral of angular momentum of orbital-like character and c) a surface integral. It was found to be essential [3, 4] to take account of the surface integral in order to understand the apparent paradox, from (1), that a plane wave appears to carry no angular momentum in the direction of propagation while experiment [5] shows that it does.

This paper deals with an expression for the "bound" angular momentum $\mathbf{J}_b$ associated with the presence of electric charge density $\rho(\mathbf{x}, t)$ that was also obtained in [2]. It is given by

$$\mathbf{J}_b(t) = \frac{1}{c} \int d^3 x \, \rho(\mathbf{x},t) \int \frac{d^3 y}{4\pi} \mathbf{y} \times [\nabla_x \times \frac{\mathbf{B}(\mathbf{y},t)}{|\mathbf{x}-\mathbf{y}|}] \qquad (2)$$

where $\nabla_x$ is the gradient operator with respect to $\mathbf{x}$. We note that if the linear coordinate vector $\mathbf{y}$ is replaced by $\mathbf{x}$ in the integral (2) above, as shown immediately following in (3), we get

$$\mathbf{J}_b' = \frac{1}{c}\int d^3x \, \rho(\mathbf{x},t) \int \frac{d^3y}{4\pi} \mathbf{x} \times [\nabla_x \times \frac{\mathbf{B}(\mathbf{y},t)}{|\mathbf{x}-\mathbf{y}|}] \quad , \quad (3)$$

and this integral becomes the algebraically simpler expression

$$\mathbf{J}_b' = \frac{1}{c}\int d^3x \, \rho(\mathbf{x},t) \mathbf{x} \times \mathbf{A}_t(\mathbf{x},t) \quad (4)$$

where the transverse component of the vector potential $\mathbf{A}_t$ is given by [6, 7]

$$\mathbf{A}_t(\mathbf{x},t) = \nabla_x \times \int d^3y \frac{\mathbf{B}(\mathbf{y},t)}{4\pi|\mathbf{x}-\mathbf{y}|} \quad . \quad (5)$$

Another expression for the bound angular momentum

$$\mathbf{J}_b' = \frac{q}{c} \mathbf{x} \times \mathbf{A}_t(\mathbf{x},t) \quad (6)$$

was obtained by Cohen-Tannoudji *et al.* [8] page 46 Equ. (7) using a different method. They expressed their result in terms of a discrete charge distribution rather than the continuous charge distribution $\rho(\mathbf{x})$. Equation (4) is the continuum generalisation of equation (6). These authors also used the abstract quantity $\mathbf{A}_t$ for the transverse vector potential. They did not provide the explicit expression for the transverse vector potential given by (5). Eq. (4) has been used elsewhere [9] to show that the electromagnetic field makes zero contribution to the angular momentum of the physical electron described by the Lagrangian of quantum electrodynamics.

Equation (2) will be mathematically equivalent to equations (3) and (4) and so $\mathbf{J}_b = \mathbf{J}_b'$ only if the following vector integral $\mathbf{I}(\mathbf{x}) \propto (\mathbf{J}_b - \mathbf{J}_b')$ vanishes

$$\mathbf{I}(\mathbf{x}) = \int d^3y (\mathbf{x}-\mathbf{y}) \times [\nabla_x \times \frac{\mathbf{B}(\mathbf{y},t)}{|\mathbf{x}-\mathbf{y}|}] \quad . \quad (7)$$

In the remainder of this paper we show that the vector integral $\mathbf{I}(\mathbf{x})$ does indeed vanish and so (2) and (4) are equivalent forms of expressing the bound angular momentum of the electromagnetic field.

By using the vector identity

$$\nabla_x \times \frac{\mathbf{B}(\mathbf{y},t)}{|\mathbf{x}-\mathbf{y}|} = \nabla_x \frac{1}{|\mathbf{x}-\mathbf{y}|} \times \mathbf{B}(\mathbf{y},t) \quad (8)$$

we can express the integrand of (7) as a triple vector product

*Bound angular momentum of the electromagnetic field*   A.M.Stewart

$$\mathbf{I}(\mathbf{x}) = \int d^3y (\mathbf{x} - \mathbf{y}) \times [\nabla_x \frac{1}{|\mathbf{x} - \mathbf{y}|} \times \mathbf{B}(\mathbf{y},t)] \qquad (9)$$

which may be expanded to give two terms $\mathbf{I}(\mathbf{x}) = \mathbf{I}_1 + \mathbf{I}_2$ where

$$\mathbf{I}_1 = -\int d^3y \{\mathbf{B}(\mathbf{y},t)[(\mathbf{x} - \mathbf{y}) \cdot \nabla_x \frac{1}{|\mathbf{x} - \mathbf{y}|}]\} \qquad . \qquad (10)$$

By multiplying out the scalar product, this comes to

$$\mathbf{I}_1 = \int d^3y \frac{\mathbf{B}(\mathbf{y},t)}{|\mathbf{x} - \mathbf{y}|} \qquad . \qquad (11)$$

The second term of the expansion of the integral (9) is

$$\mathbf{I}_2 = \int d^3y [\mathbf{B}(\mathbf{y},t) \cdot (\mathbf{x} - \mathbf{y})] \nabla_x \frac{1}{|\mathbf{x} - \mathbf{y}|} \qquad . \qquad (12)$$

To evaluate $\mathbf{I}_2$ we consider the identity

$$\frac{\partial}{\partial y^m}[\frac{(x^i - y^i)B^m}{|\mathbf{x} - \mathbf{y}|}] = (x^i - y^i)B^m \frac{\partial}{\partial y^m}(\frac{1}{|\mathbf{x} - \mathbf{y}|}) + \frac{x^i - y^i}{|\mathbf{x} - \mathbf{y}|} \frac{\partial B^m}{\partial y^m} - \frac{B^m}{|\mathbf{x} - \mathbf{y}|} \delta_{i,m} \qquad . \qquad (13)$$

If this identity is integrated over $y^m$ from minus infinity to plus infinity the left hand side vanishes because the integrand is zero at those values and we obtain

$$\int_{-\infty}^{\infty} dy^m (x^i - y^i)B^m \frac{\partial}{\partial y^m}(\frac{1}{|\mathbf{x} - \mathbf{y}|}) + \int_{-\infty}^{\infty} dy^m \frac{x^i - y^i}{|\mathbf{x} - \mathbf{y}|} \frac{\partial B^m}{\partial y^m} - \int_{-\infty}^{\infty} dy^m \frac{B^m}{|\mathbf{x} - \mathbf{y}|} \delta_{i,m} = 0 \qquad . \qquad (14)$$

If we now integrate over the other two components of $\mathbf{y}$ the integral becomes a volume integral of $d^3y$ over all space and, summing over $m$, the second term vanishes from $\nabla_y \cdot \mathbf{B}(\mathbf{y}) = 0$.

We get
$$\sum_m \int d^3y (x^i - y^i)B^m \frac{\partial}{\partial y^m}(\frac{1}{|\mathbf{x} - \mathbf{y}|}) - \sum_m \int d^3y \frac{B^m}{|\mathbf{x} - \mathbf{y}|} \delta_{i,m} = 0 \qquad . \qquad (15)$$

or
$$\int d^3y (x^i - y^i) \mathbf{B} \cdot \nabla_x (\frac{1}{|\mathbf{x} - \mathbf{y}|}) = -\int d^3y \frac{B^i}{|\mathbf{x} - \mathbf{y}|} \qquad . \qquad (16)$$

So, writing out the gradient explicitly,

$$\mathbf{I}_2 = -\int d^3y \frac{\mathbf{B}(\mathbf{y},t)}{|\mathbf{x} - \mathbf{y}|} \qquad . \qquad (17)$$

Equation (17) cancels (11) so it is proved that the integral $\mathbf{I}(\mathbf{x})$ vanishes, $\mathbf{J}_b = \mathbf{J}_b'$, and the two mathematical forms (2) and (4) for the "bound" angular momentum are equivalent.

The bound linear momentum $\mathbf{P}_b$ comes to

$$\mathbf{P}_b = \frac{1}{c}\int d^3x\, \rho(\mathbf{x},t)\mathbf{A}_t(\mathbf{x},t) \qquad . \qquad (18)$$

This equation was printed incorrectly in equation (34) of [2]. Also, the second part of equation (4) of [2] should be given a minus sign to read

$$\mathbf{F}(\mathbf{x},t) = -\frac{\partial}{\partial t}\int d^3y\, \frac{\mathbf{B}(\mathbf{y},t)}{4\pi c\,|\mathbf{x}-\mathbf{y}|} \qquad , \qquad (19)$$

and the equation in the second line of text above equation (6) of that paper [2] should read $\nabla\cdot\mathbf{B} = 0$ not $\nabla\times\mathbf{B} = 0$.